# Effects of ozone post deposition treatment on interfacial and electrical characteristics of atomic-layer-deposited $Al_2O_3$ and $HfO_2$ films on GaSb substrates

Lianfeng Zhao, Zhen Tan, Jing Wang, and Jun Xu[*]

Tsinghua National Laboratory for Information Science and Technology, Institute of Microelectronics, Tsinghua University, Beijing 100084, People's Republic of China

**Abstract**

Atomic-layer-deposited $Al_2O_3$ and $HfO_2$ films on GaSb substrates were treated by *in-situ* ozone post deposition treatment (PDT). The effects of ozone PDT on the interfacial and electrical properties of $Al_2O_3$ and $HfO_2$ gate dielectric films on GaSb substrates were investigated carefully. It is found that the dielectric quality and the interfacial properties of the $Al_2O_3$ and $HfO_2$ films are improved by ozone PDT. After *in-situ* ozone PDT for 5 min, the $Al_2O_3$ and $HfO_2$ films on GaSb substrates exhibit improved electrical and interfacial properties, such as reduced frequency dispersion, gate leakage current, border traps and interface traps. Interface trap density is reduced by ~24% for the $Al_2O_3$/GaSb stacks and ~27% for the $HfO_2$/GaSb stacks. *In-situ* ozone PDT is proved to be a promising technique in improving the quality of high-k gate stacks on GaSb substrates.

[*]Corresponding author. Tel.: +86-10-62794752; Fax:+86-10-62771130.

E-mail address: junxu@tsinghua.edu.cn (J. Xu).

## 1. Introduction

As the end of silicon complementary metal-oxide-semiconductor (CMOS) device scaling is approaching, III-V semiconductor materials have attracted much attention due to their higher carrier mobility and lower effective mass [1]. Although most of the III-V compound semiconductors, such as GaAs and InAs, possess high electron mobility and saturation velocity, their hole mobility is relatively low, making them unsuitable as p-channel materials [2]. GaSb is an attractive material for p-channel application, because of its high hole mobility ($\sim 1000\ cm^2V^{-1}s^{-1}$) [3]. Up to now, GaSb devices have been intensively investigated [4,5]. However, there are many issues impeding the wide application of GaSb MOS devices, such as high interface trap density ($D_{it}$). This is mostly due to the lack of high-quality, thermodynamically stable native oxide with low interface trap density. Although adopting high-k dielectrics such as $Al_2O_3$ and $HfO_2$ have benefits in device scaling, high-k dielectrics deposited directly on GaSb surfaces generally exhibit high $D_{it}$. Many methods have been proposed to improve the interfacial properties of high-mobility MOS devices beyond silicon, such as neutralized sulfur passivation [6], plasma treatment [7,8], insertion of an interfacial layer [9,10], and the use of novel dielectrics [11,12]. However, ozone treatment has rarely been investigated, and previously reported results were mainly focused on surface oxidation [13]. In this paper, the effects of *in-situ* ozone post deposition treatment (PDT) on high-k dielectrics on GaSb substrates are investigated. It is found that great improvements in interfacial and electrical properties are achieved by this technique.

## 2. Experimental details

N-type Te-doped (100)-oriented GaSb wafers with a doping concentration of $\sim 10^{17}$ cm$^{-3}$ were used as substrates in this work. The substrates were first degreased by being sequentially immersed for 5 min each in acetone, ethanol, and isopropanol, and then cleaned by 9% HCl aqueous solution for 1 min. Sulfur passivation was then performed by immersing the substrates into 15% $(NH_4)_2S$ aqueous solutions for 15 min at room temperature. After that, ~8 nm $Al_2O_3$ or $HfO_2$ dielectrics were deposited by atomic layer deposition (ALD) at 200 °C using Beneq TFS 200 ALD system. Trimethylaluminum (TMA) and water were used as precursors for $Al_2O_3$ deposition; and Tetrakis(ethylmethylamino)hafnium (TEMAH) and water were used as precursors for $HfO_2$ deposition. Then, *in-situ* ozone PDT was performed for 5 min for different sample groups at 200 °C. The average ozone pressure in the ALD Reactor is controlled at ~8 mbar. For measuring electrical properties, Al was evaporated and patterned to form gate contacts and back metal contacts of Ti/Au were also deposited. Control samples without ozone treatment were fabricated for comparison. Capacitance-voltage (C-V), conductance-voltage (G-V), and current-voltage (I-V) characteristics were recorded using an Agilent B1500A semiconductor device analyzer and Cascade Summit 11000 AP probe system. High resolution transmission electron microscope (HRTEM) measurements were performed using a Tecnai G2 F20 S-Twin TEM for physical characterization of the high-k/metal-gate stacks.

## 3. Results and discussions

Fig. 1 illustrates the high frequency C-V hysteresis characteristics of $Al_2O_3$ on GaSb substrates with and without *in-situ* ozone PDT. It clearly shows that the oxide capacitance ($C_{ox}$) in the accumulation region is similar for all samples, which indicates that the 5 min ozone PDT does not result in the formation of an additional oxide interfacial layer. The equivalent oxide thickness (EOT) for the samples with ozone PDT is calculated to be ~4.8 nm, which is consistent with the physical thickness of ~8.2 nm (confirmed by the HRTEM measurements) and the previously reported dielectric constant of ~7 for ALD deposited $Al_2O_3$ films [14]. The reduced C-V hysteresis for samples with ozone treatment suggests that border traps are reduced by ozone PDT [15]. A measure of the total effective border traps density ($\Delta N_{bt}$) can be obtained by integrating the absolute value of the difference in C-V hysteresis from reverse to forward bias direction ($\Delta C$), using the expression [15, 16]:

$$\Delta N_{bt} = \frac{1}{qA} \int |\Delta C| dV \quad (1)$$

where q is the elementary charge and A is the area of the $Al_2O_3$/GaSb stacks. Fig. 2 plots the border traps estimation curves ($\Delta C/q$ *versus* $V_{gate}$) for samples with and without ozone treatment. The extracted $\Delta N_{bt}$ of the $Al_2O_3$/GaSb stacks are reduced by 18.5%, from ~$2.48 \times 10^{12}$ cm$^{-2}$ to ~$2.02 \times 10^{12}$ cm$^{-2}$, after ozone PDT.

Normalized C-V curves with a small AC signal ranging from 1 kHz to 1 MHz were measured to compare the frequency dispersion of the samples (Fig. 3). The frequency dispersion in the accumulation region is reduced from 8.4% to 4.5% by ozone PDT, which indicates that interfacial properties are greatly improved by this technique [17]. Fig. 4(a) shows typical measured parallel $G_p/\omega$ *versus* frequency curves for different gate biases of the $Al_2O_3$/GaSb stacks with ozone PDT. The observed peak shift indicates the efficiency of the surface Fermi level movement over the energy gap [18]. $D_{it}$ distribution is extracted using the conductance method [19, 20], as shown in Fig. 4(b). After *in-situ* ozone PDT, $D_{it}$ is reduced by ~24%, which proves the ozone PDT to be a promising technique to improve the interfacial properties between high-k dielectrics and GaSb substrates.

Fig. 5 shows the gate leakage current characteristics of $Al_2O_3$ on GaSb substrates with and without *in-situ* ozone PDT. The reduced gate leakage current for samples with the ozone treatment indicates a reduction in trap-assisted current, which is consistent with the reduced border traps and interface traps [16].

Fig. 6 shows the HRTEM images of the $Al_2O_3$/GaSb stacks with and without ozone PDT. After ozone PDT, the interface between $Al_2O_3$ films and GaSb substrates becomes sharper and flatter, and the $Al_2O_3$ films are more uniform, which indicates the improved quality of $Al_2O_3$ dielectrics and interfacial condition. This is consistent with the reduced gate leakage current. In this work, no thermal annealing was performed because of the low thermal budget of Antimonides, leading to defects and vacancies in the dielectrics. Oxygen provided by ozone PDT can go through the dielectrics and reach the interface, reducing the defects and vacancies in the dielectrics and near the interface, thereby improving the properties of dielectrics and interfaces.

The effects of ozone PDT on the $HfO_2$/GaSb stacks were also investigated. Fig. 7 shows the gate leakage current characteristics of the $HfO_2$/GaSb stacks with and without ozone PDT. Compared with the samples without ozone treatment, an obvious reduction in gate leakage current is found for the $HfO_2$/GaSb stacks with the ozone treatment. This reduction is much more significant than that of the $Al_2O_3$/GaSb stacks. As shown in the inset of Fig. 7, high frequency C-V characteristics at 1 MHz of these samples are similar, indicating a similar EOT for all the samples. The EOT of these samples is calculated to be ~3.5 nm, which is not increased by the ozone treatment. Since EOT is similar for all the samples, the reduced gate leakage might be caused by a reduction in border traps and interface traps [16], which is confirmed by C-V and G-V measurements below.

C-V hysteresis characteristics were measured to investigate the border traps. As shown in Fig. 8, a large hysteresis is observed for samples without ozone treatment, which indicates large border traps [15]. C-V hysteresis for samples with ozone treatment is reduced significantly, suggesting that border traps are reduced considerably. Fig. 9 demonstrates the border traps estimation curves of the $HfO_2$/GaSb stacks with and without ozone PDT. Before ozone treatment, $\Delta N_{bt}$ of the HfO$_2$/GaSb stacks is as large as ~$6.94 \times 10^{12}$ cm$^{-2}$, whereas after ozone treatment, $\Delta N_{bt}$ is significantly reduced to ~$2.29 \times 10^{12}$ cm$^{-2}$. This reduction is much more significant than that of the Al$_2$O$_3$/GaSb stacks.

Interface properties were investigated by measuring G-V characteristics of the samples with and without ozone treatment. Fig. 10 (a) plots the measured parallel $G_p/\omega$ *versus* frequency curves of the HfO$_2$/GaSb stacks with ozone PDT. The shift of the curve peaks for different gate biases suggests that Fermi level pinning is eliminated [18]. $D_{it}$ distribution in the band-gap was determined using the conductance method [19, 20], as plotted in Fig. 10 (b). $D_{it}$ of the HfO$_2$/GaSb stacks with ozone PDT is reduced by ~27% compared with those without ozone treatment, which is also more significant than that of the Al$_2$O$_3$/GaSb stacks.

Ozone treatment is shown to be more effective in improving the performance of the HfO$_2$/GaSb stacks than that of the Al$_2$O$_3$/GaSb stacks. This is because the properties of the as-deposited Al$_2$O$_3$ films are better than that of the HfO$_2$ films, due to the following two reasons. First, the deposition temperature is not optimal for HfO$_2$ films. Because of the low thermal budget of Antimonides, 200 °C is set as the HfO$_2$ deposition temperature. However, previous research has shown that 200 °C is not an optimal temperature for HfO$_2$ deposition, and that film density increases as deposition temperature is further increased [21]. Consequently, there are more defects in the HfO$_2$ films, which is consistent with the larger border traps density of the HfO$_2$/GaSb stacks than that of the Al$_2$O$_3$/GaSb stacks. Second, Al$_2$O$_3$ films generally obtain better interface properties on III-V substrates, such as InGaAs and GaSb, than that of the HfO$_2$ films [22, 23]. As a result, before ozone treatment, the interface trap density of the HfO$_2$/GaSb stacks is higher than that of the Al$_2$O$_3$/GaSb stacks. After ozone treatment, border traps and interface traps are reduced for both the Al$_2$O$_3$ and the HfO$_2$ films on GaSb substrates, and the difference in $\Delta N_{bt}$ and $D_{it}$ between HfO$_2$ and Al$_2$O$_3$ films is smaller than that of the as-deposited films.

## 4. Conclusions

We have systematically studied the effects of ozone PDT on the electrical and interfacial properties of Al$_2$O$_3$ and HfO$_2$ dielectrics on GaSb substrates. Great improvements in frequency dispersion, gate leakage current, border traps and interface traps are achieved by the *in-situ* ozone PDT. This is because defects at the interface and in the dielectrics are greatly reduced. Consequently, *in-situ* ozone PDT is proved to be a promising technique in improving the performance of modern III-V MOS devices. Furthermore, the improved properties of the Al$_2$O$_3$ and HfO$_2$ films by ozone treatment indicate that ozone treatment is a promising technique in improving the quality of high-k oxide films deposited at low temperatures without further annealing.

**Acknowledgments**

This work was supported in part by the State Key Development Program for Basic Research of China (No. 2011CBA00602), and the National Science and Technology Major Project (No. 2011ZX02708-002).

**Figure Captions**

Fig. 1. High frequency C-V hysteresis curves at 1 MHz of the $Al_2O_3$/GaSb stacks with and without ozone PDT.

Fig. 2. Border traps estimation curves of the $Al_2O_3$/GaSb stacks with and without ozone PDT.

Fig. 3. Normalized multi-frequency C-V characteristics of the $Al_2O_3$/GaSb stacks (a) without ozone PDT and (b) with ozone PDT.

Fig. 4. (a) Typical measured parallel $G_p/\omega$ versus frequency curves for different gate biases of the $Al_2O_3$/GaSb stacks with ozone PDT. (b) $D_{it}$ distribution of the $Al_2O_3$/GaSb stacks with and without ozone PDT.

Fig. 5. Gate leakage current characteristics of the $Al_2O_3$/GaSb stacks with and without ozone PDT.

Fig. 6. HRTEM images of the $Al_2O_3$/GaSb stacks (a) without ozone PDT and (b) with ozone PDT.

Fig. 7. Gate leakage current characteristics of the $HfO_2$/GaSb stacks with and without ozone PDT. The inset shows high frequency C-V characteristics of the $HfO_2$/GaSb stacks at 1 MHz.

Fig. 8. C-V hysteresis characteristics at 1 MHz of the $HfO_2$/GaSb stacks with and without ozone PDT.

Fig. 9. Border traps estimation curves of the $HfO_2$/GaSb stacks with and without ozone PDT.

Fig. 10. (a) Typical measured parallel $G_p/\omega$ versus frequency curves of the $HfO_2$/GaSb stacks with ozone PDT. (b) $D_{it}$ distribution of the $HfO_2$/GaSb stacks with and without ozone PDT.

Figures

Fig. 1

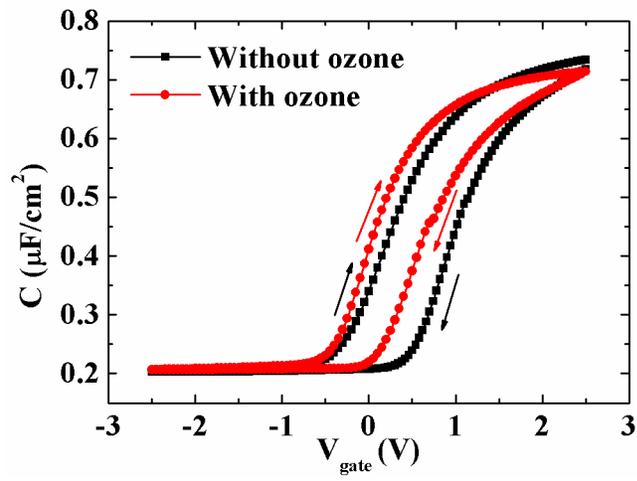

Figures (continued)

Fig. 2

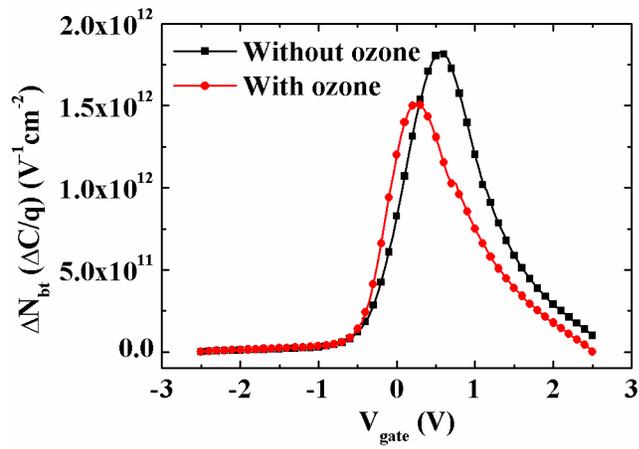

Figures (continued)

Fig. 3

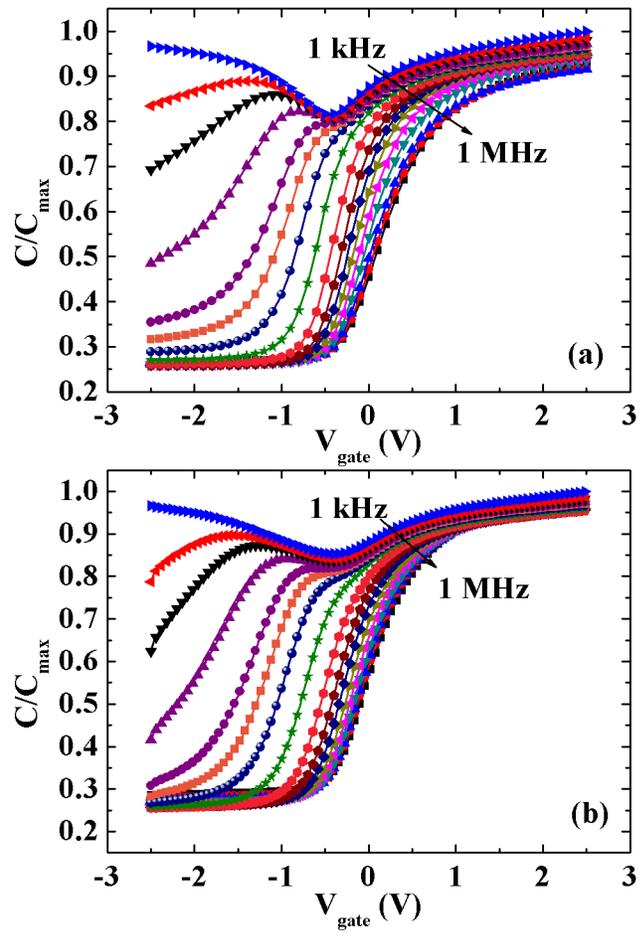

Figures (continued)

Fig. 4

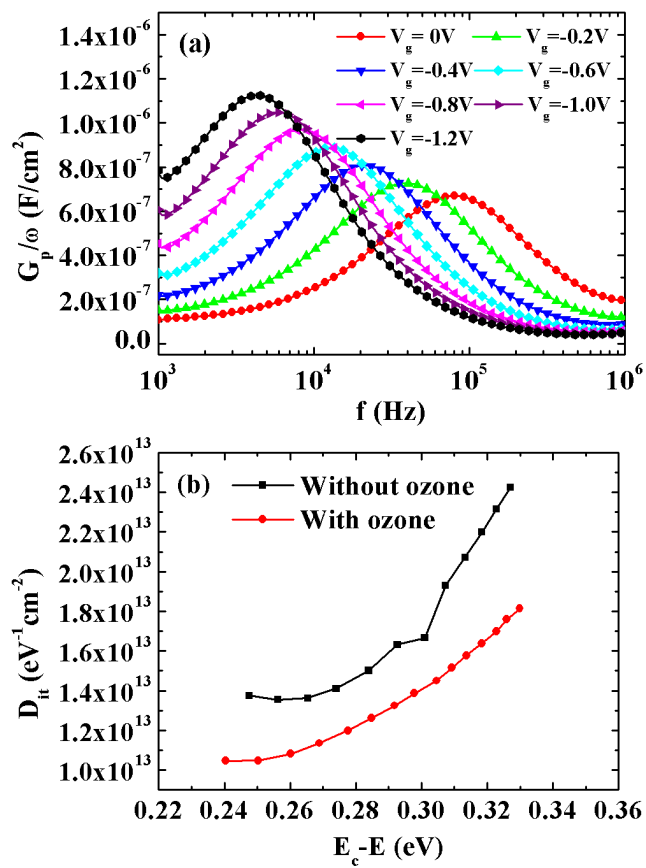

Figures (continued)

Fig. 5

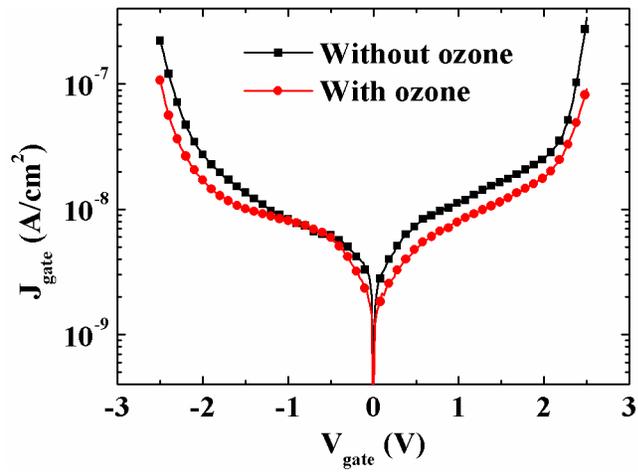

Figures (continued)

Fig. 6

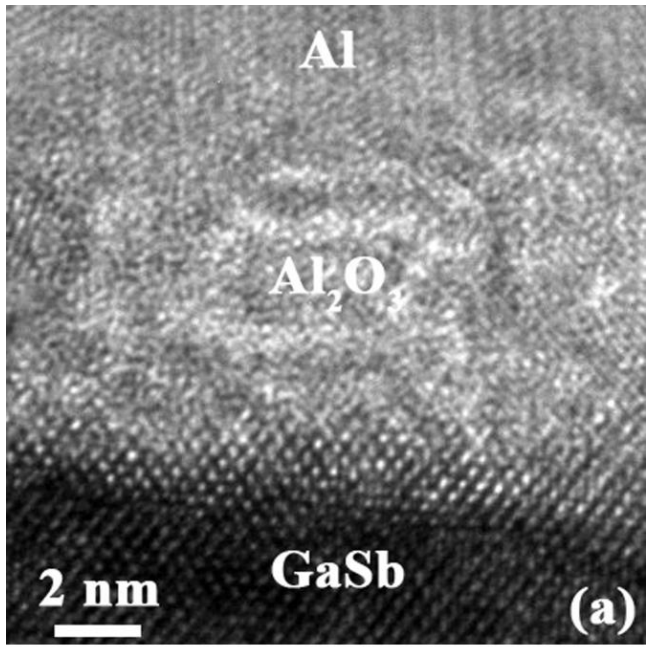

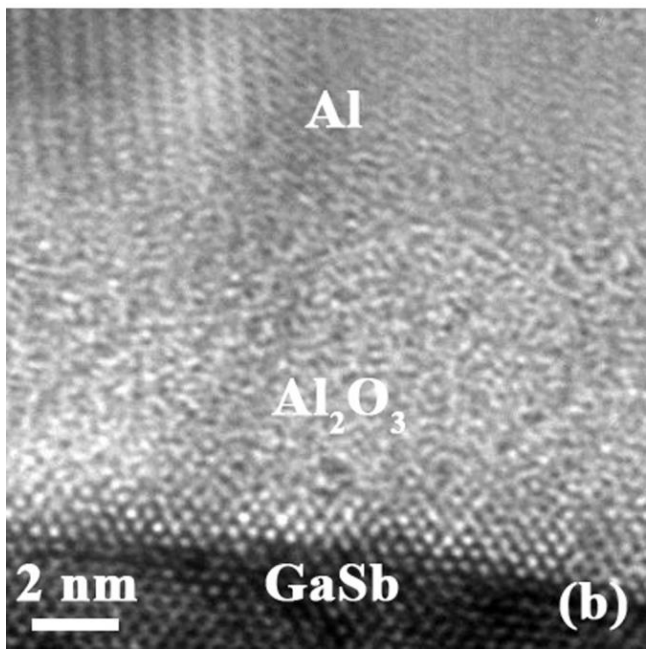

Figures (continued)

Fig. 7

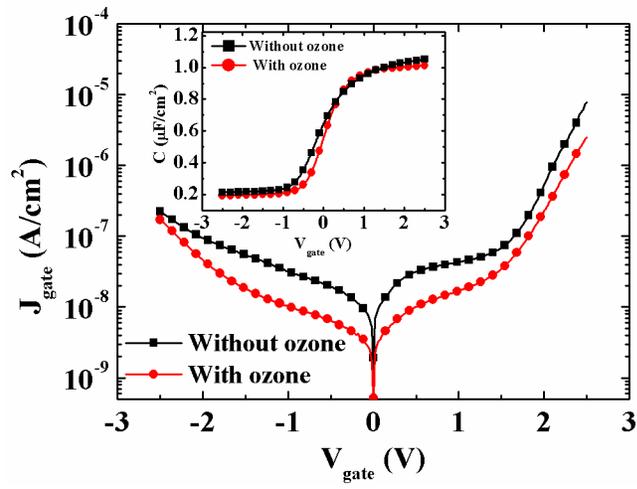

Figures (continued)

Fig. 8

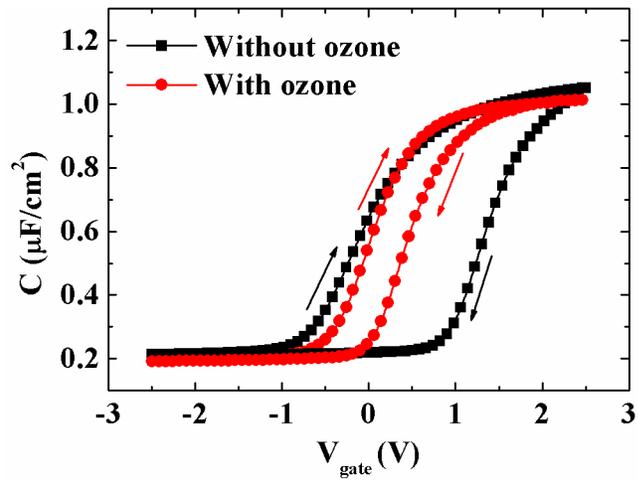

Figures (continued)

Fig. 9

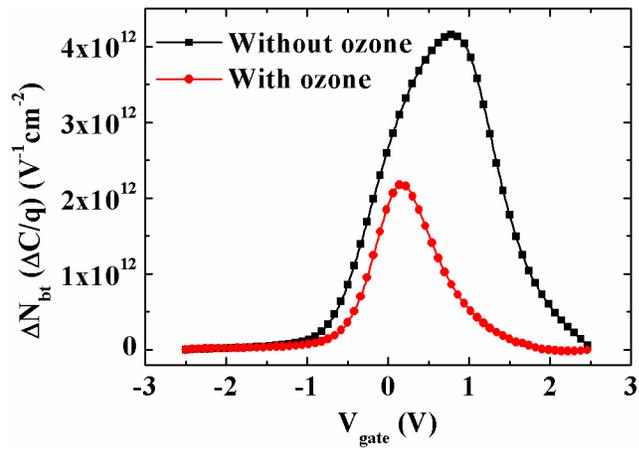

Figures (continued)

Fig. 10

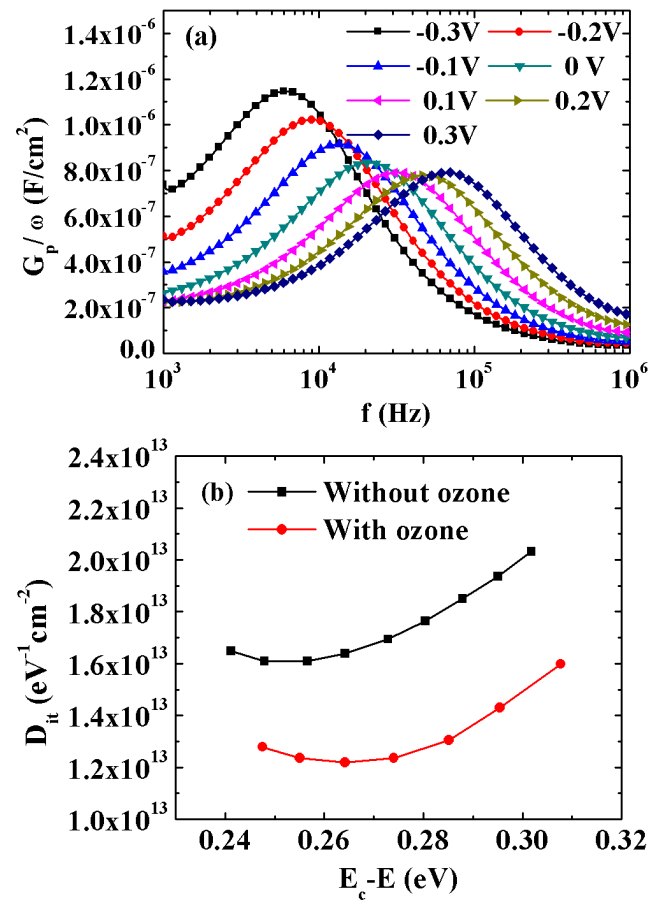